\documentclass[conference]{IEEEtran}
\IEEEoverridecommandlockouts
\usepackage{amsmath,amssymb,amsfonts}
\usepackage{algorithmic}
\usepackage{graphicx}
\usepackage{textcomp}
\usepackage{xcolor}
\usepackage{hyperref}
\usepackage[backend=biber,style=alphabetic,sorting=ynt]{biblatex}
\def\BibTeX{{\rm B\kern-.05em{\sc i\kern-.025em b}\kern-.08em
    T\kern-.1667em\lower.7ex\hbox{E}\kern-.125emX}}
\begin{document}

\title{Avatar Fusion Karaoke: Research and development on multi-user music play VR experience in the metaverse}

\author{\IEEEauthorblockN{1\textsuperscript{st} Alexandre BERTHAULT}
\IEEEauthorblockA{\textit{GREE VR Studio Laboratory} \\
\textit{REALITY, Inc.}\\
Tokyo, Japan \\
apro.berthault@gmail.com}
\and
\IEEEauthorblockN{2\textsuperscript{nd} Takuma KATO}
\IEEEauthorblockA{\textit{GREE VR Studio Laboratory} \\
\textit{REALITY, Inc.}\\
Tokyo, Japan \\
takuma.katou42@gmail.com}
\and
\IEEEauthorblockN{3\textsuperscript{rd} Akihiko SHIRAI}
\IEEEauthorblockA{\textit{GREE VR Studio Laboratory} \\
\textit{REALITY, Inc.}\\
Tokyo, Japan \\
shirai@mail.com}
}

\maketitle

\begin{abstract}
This paper contributes to building a standard process of research and development (R\&D) for new user experiences (UX) in metaverse services. We tested this R\&D process on a new UX proof of concept (PoC) for Meta Quest head-mounted display (HMDs) consisting of a school-life karaoke experience with the hypothesis that it is possible to design the avatars with only the necessary functions and rendering costs. The school life metaverse is a relevant subject for discovering issues and problems in this type of simultaneous connection. 
To qualitatively evaluate the potential of a multi-person metaverse experience, this study investigated subjects where each avatar requires expressive skills. While avatar play experiences feature artistic expressions, such as dancing, playing musical instruments, and drawing, and these can be used to evaluate their operability and expressive capabilities qualitatively, the Quest's tracking capabilities are insufficient for full-body performance and graphical art expression. 
Considering such hardware limitations, this study evaluated the Quest, focusing primarily on UX simplicity using AI Fusion techniques and expressiveness in instrumental scenes played by approximately four avatars.
To achieve our PoC, we first researched the limitations and challenges of such applications. 
In Meta Quest 1 and Meta Quest 2, the benchmark for simultaneous avatar rendering was 10 for Quest 1 and 22 for the Quest 2. 
The appropriate interpersonal distance (IPD) should be designed with culture in mind. 
We assume that various avatars communicate at a maximum distance of 1–2 meters while generating collisions for individual avatars and avoiding overlap. Considering the interference in stereophonic sound, we consider that 5–11 avatars can be displayed simultaneously while maintaining quality communication.
Through our standard R\&D process, we assessed the production needs for our PoC and obtained multiple ways to retrieve feedback. 
A notable effect was observed when other users entered the same session. When users waved and called out to one another (saluted), their participation time increased by 3 minutes (i.e., they played more than one piece of music). Most users did not perceive the AI-only piano-playing avatar as human (i.e., they perceived it as a non-player character [NPC] in the game).
This research reported methods for multi-user metaverse communication and its supporting technologies, such as head-mounted devices and their graphics performance, special interaction techniques, and complementary tools and the importance of PoC development, its evaluation, and its iterations. The result is remarkable for further research; these expressive technologies in a multi-user context are directly related to the quality of communication within the metaverse and the value of the user-generated content (UGC) produced there.

\end{abstract}


\section{Introduction}
\begin{figure*}[ht]
\centerline{\includegraphics[width=1\linewidth]{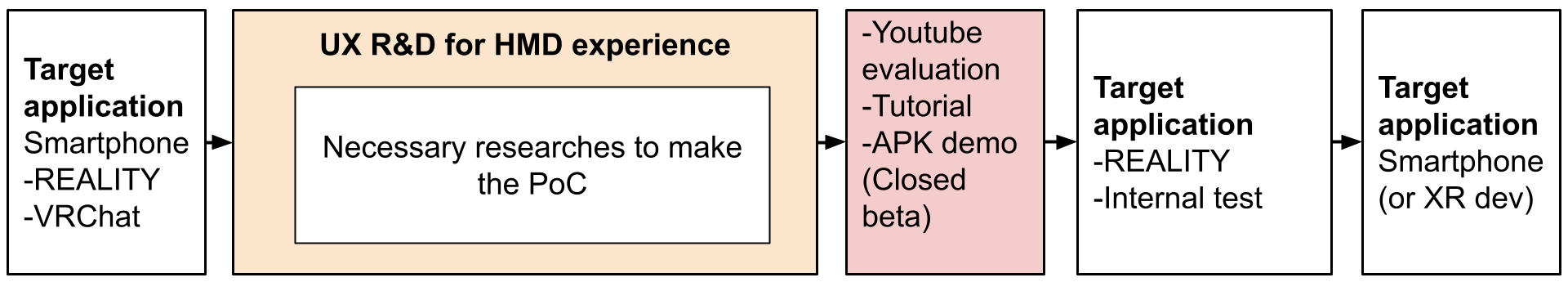}}
\caption{R\&D procedure. This paper focuses on completing UX R\&D for the HMD experience.}
\label{fig-1}
\end{figure*}
\subsection{Context and motivation}
The importance of the simultaneity of 3D space and UGC is mentioned in the modern definition of the metaverse (Matthiu Ball)\cite{ball2022metaverse}.
The underlying motivation for this study was based on the idea that we need to create an atavistic experience in existing or future HMD devices from the avatar-driven social networking service (SNS) that is expanding on smartphones today.
People have many expectations and fantasies about the metaverse; therefore, it must first be defined and scoped. A 3D-avatar social networking service, the metaverse has been the focus of much attention in recent years, with ZEPETO, REALITY, and other services growing rapidly. 
Assuming an inevitable future when extended reality (XR) devices, such as HMDs and augumented reality (AR) glasses are widely adopted by current smartphone users, various related topics will require research. In addition to human-computer interaction (HCI) research using current virtual reality ready devices, such as PCs and stand-alone HMDs, R\&D will be needed on the avatar UX that contemporary, casual smartphone users would naturally find attractive as an extended-reality social network service (XRSNS).

\subsection{Avatar expression and related works}
The user avatar is another crucial subject. Users  such as VTubers and non-fungible token (NFT) artists employ varied identities as avatars; avatars are also used as Virtual Beings in distributor culture and as icons on many platforms and social media. One trend maintains that our 3D avatars should be photorealistic, but this depends on the UX and functions of the service, the rendering performance, and the relationship (culture) between one user and another. 
In our research, simulating reality is not necessarily the goal, as making realistic avatars faces the challenge of the uncanny valley.
Although many avatars attempt to overcome the uncanny valley, it is necessary not only to improve the quality and elements of CG rendering, such as reflection and lighting, but also to remain attentive to eye tracking and reaction speed. 
Research on Telexistence’s robots, such as the Telenoid developed by Ishiguro et al. (https://robots.ieee.org/robots/telenoid/), propose that it is preferable to build only abstract facial elements in a general-purpose avatar. 
However, when many metaverse services and XR devices are released, such discussions from a R\&D perspective may be meaningless.
The strongest reason for the debate between photorealistic and abstract avatars is the availability of consumer hardware, such as the Meta Quest. 
In the past, PCVR-only metaverse services, or legacy environments in the SecondLife era, only needed to create a comfortable UX environment; current metaverse services should  focus on more generic users who will not consider the specifications. 
Setting up standard hardware as represented by smartphones and the Quest, and developing a comfortable UX evaluation method for quality assurance in the metaverse can increase the number of users.

REALITY is a smartphone application that focuses on manga and anime aesthetics and streaming features to create a world where teenagers, Generation Z, and manga and anime enthusiasts can enjoy chatting with friends, giving gifts in a manga and anime world, and wearing anime and manga–themed clothing. 
What style and quality should the renderings in such a world have? A photorealistic style would not be suitable. Abstract avatars, or ones that significantly differ from one’s actual physical features, would be more appropriate. Users could also choose their avatars to suit a manga or sci-fi aesthetic. Rendering methods based on a cartoonish style and unlit lighting can be used to avoid the uncanny valley. 
As several studies paradoxically prove, it is not necessary to be photorealistic to match realistic attributes.
AgileAvatar takes feature vectors from a live-action photo and applies them to the placement parameters of encoded avatar elements (like the Nintendo Mii). Although this method sets reality as the Ground Truth, it approximates a rendering of a one-shot cartoonish image, so there may be a change of purpose. 
Users who want an avatar of themselves rendered like a 2D photo that they can use to interface with society are only a subset of all users.
Instead, users want an ideal picture of themselves, so a process similar to using a beauty filter is more in demand.

\subsection{Objectives and challenges}
This paper proposes a standard R\&D process in the metaverse as a hypothesis. First, there is the overall picture of the new UX that metaverse users seek and the functions required of avatars that cannot be fixed. Since these diverge infinitely, they have different characteristics from game content. 
This study defines user characteristics as ``a proposal of attractive engagement that teenagers who like anime and manga will experience using XR devices in the near future.'' 
Therefore, the challenge will be to research and achieve a PoC that can answer these user characteristics. 
R\&D should develop a PoC that facilitates evaluating a practical application as a commercial service through quality and user testing conducted on the platform. 
Figure 1 illustrates the process that identifies the technology groups and R\&D problems to be solved to complete the technical PoC and prepare for evaluation through internal and user testing.

\section{Related Works: Multiplayer engagement in metaverse}
In our research on new features for our target application, we track recent advances in the metaverse state of the art and overall mixed reality experiences. 
We present some prior research on metaverse UX evaluations. 
In these studies, we found some significant elements on the value of metaverse UX development.
Empirical Research on the Metaverse User Experience of Digital Natives (Lee and Gu)\cite{Sustainability} evaluated ZEPETO and other platforms horizontally. 
User-friendly details were added, and users were satisfied and experienced high usability on the rapidly developing metaverse platform. 
The findings from this study's proposed method led to a practical research outcome that identified how UX, as an important element of metaverse platform usage optimization, could be convenient and effective. 
Additionally, it can be sustained when the number of convergence platforms, digital services, and their compatible IT devices significantly increases. 
This pioneered empirical research evaluating the usability of the worldwide metaverse platform from digital natives' point of view and offered some suggestions for future research. 
The methodological limitations caused by the small number of interview surveys can be addressed with additional follow-up research using larger participant samples more diverse perspectives. 
In addition, depending on the metaverse's industrial and technological maturity, a richer level of understanding will be possible if a large-scale user survey is conducted in parallel and comparatively analyzed with the results. 
Finally, a usability evaluation would still have to be structured and assessed based on fully considered multidimensional factors originating from the complexity of the participants’ usage circumstances. 
It is expected that in-depth research tasks for the upcoming extended real world study can be derived if additional studies by various user groups are conducted based on this study.
The Effects of User Experience-Based Design Innovativeness on User Metaverse Platform Channel Relationships in South Korea (Jeon)\cite{UXBDI} focused on user experience-based design innovativeness (UXBDI). 
The author concluded that many companies prefer an interaction strategy before, during, and after use to enhance metaverse platforms’ engagement, including all emotional, cognitive, and physical reactions. 
This strategy suggests that metaverse platforms are characterized by functional, symbolic, and esthetic dimensions that jointly determine users’ responses. 
They continuously provide new experiences and capabilities. Marketers can capitalize on UXBDI's dynamic effects over time by planning and implementing incentive promotions with users they hope will commit to metaverse platforms.
Generic experience, which can be copied from current real experience, appears to be in demand as a common expectation in mass marketing. 
A review of current metaverse-related computer entertainment and communication services and research shows, that UX is designed not only as a function of user play but also of communication. 
The core users who attract other users will be more creative and communicative. 
This creates a challenge for service operators, who must design rich, high-quality experiences, places and functions for users (creators) to harmoniously with other users. 
Artificial Intelligence for the Metaverse: A Survey (Huynh-The et al., 2022)\cite{MetaverseSurvey} concluded that AI assistants powered by conversational AI could serve many specific purposes in multilevel philosophical conversations to enhance user interactive experience. Conversational AI is a set of AI technologies (e.g., automatic speech recognition, language processing, advanced dialogue management, and machine learning) that offer humanlike interactions in the metaverse by recognizing speech and text, including understanding intention, deciphering languages, and responding by imitating human conversations over voice modality. 
Most current metaverse projects limit users to exploring, owning, and customizing things in the virtual world. 
In the future, users will be able to create hyperreal objects and content easily and quickly with AI's assistance. 
Various hyperreal objects (e.g., faces, bodies, plants, animals, vehicles, buildings, and other inanimate objects) can be remixed endlessly to design unique experiences and inspire creation. Accordingly, combining VR and AI-based content generation can immerse a user completely in alternate realities. 
AI tools to compensate for hardware limitations to simulate realistic animation. In this context, AI tools should be affordable and offer user-friendly interfaces. 
Further, ethical issues relating to the user-generated metaverse need to be seriously examined, guidelines and policies should and threats to individuals and societies when users synthesize hyperreal media content.

Following our standard protocol, our team focused on music, one of the oldest activities in human history and one not directly related to survival, as well as on musical instruments and singing. 
Musical performances using avatars in the metaverse have attracted attention, such as Travis Scott's Fortnite Live, but not everyone can play music or sing during these events. 
In this study, wetest our standard R\&D process and research the concept and technical implementation of avatar karaoke, which creates a diverse UGC for song performance and singing as individual expression assisted by AI.
In VR Karaoke Using Expressive 3D Avatars (JaeHyeok Choi et al.)\cite{VRKaraoke}, the authors created a VR platform that allows users to enjoy karaoke at home or in any private setting as a response to public karaoke venues being closed due to Covid-19.
Their platform is designed to analyze and score users’ singing performances in real time and incorporate emotional 3D avatar reactions to make them feel as if they were in a karaoke room with other people, thus helping people cope with the stress of Covid-19 isolation. 
This platform assesses the user’s singing based on the beat, notes, and lyrics and allows users to develop their singing skills by delivering real-time feedback via a scoring algorithm, in addition to creating an entertaining environment. 
It has its own analysis tool and uses both official and unofficial songs (which are unavailable on karaoke machines). 
Karaoke singers would like to copy and evaluate existing karaoke in the metaverse, but these users would not be the only ones. 
Because a lot of music is not played by one person. Karaoke is also meant to be orchestral, and will require orchestration on a metaverse.
 We focused on other musical experiences, such as small band performances. The goal is not only to play well, but also to give a performance that appeals to everyone. We are developing a VR metaverse for air guitar, in which the guitarist is not pretending to play but is in fact performing.

\section{Performance and proximity in the metaverse}

To create a PoC for our Avatar Fusion Karaoke we first researched  the challenge of creating such applications from a performance and look dev point of view.

\subsection{Performance and avatar number}
\begin{figure}[!ht]
\centerline{\includegraphics[scale=0.25]{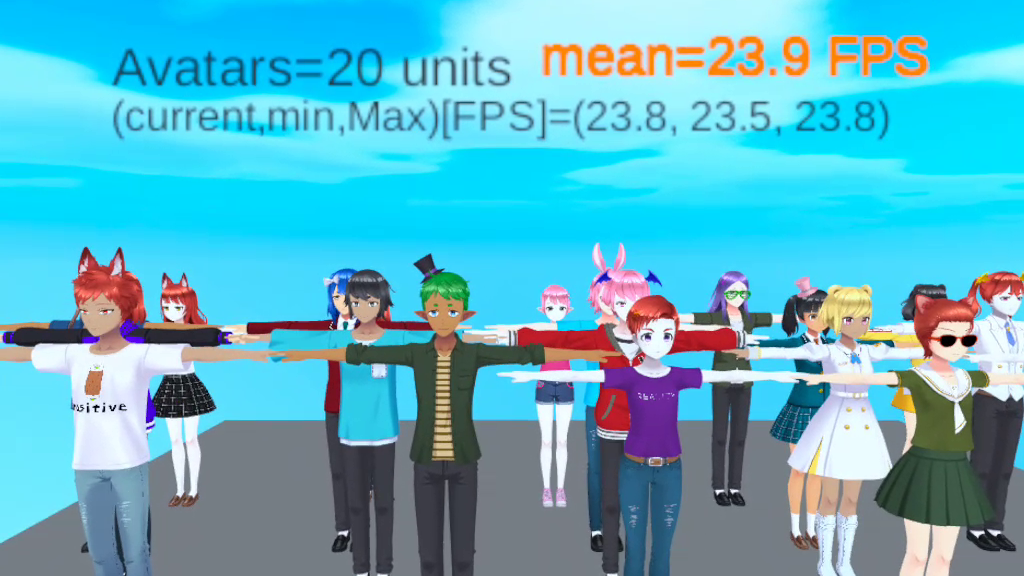}}
\caption{We created a benchmark application to control the number of avatars displayed while having a real-time monitoring interface}
\label{fig-2}
\end{figure}

When considering UX in the school-life metaverse, we wanted to create an experience as close as possible to a teenager's school experience. We researched how to make a cross-platform metaverse that used PCVR and standalone HMDs, such as Meta Quest 1 and Meta Quest 2 as possible environments. Avatars in our target application have an average polygon count of approximately 30,000 and a material count of 6~14, heavier than avatars used in general mobile applications. Figure 3 shows the loads of the avatars in Quest 1 and Quest 2. When the frames per second (FPS) drop below 60, a slight delay is perceived, and when the FPS drops below 40, the delay becomes obvious. On Quest 2, the FPS starts dropping when more than 22 avatars are rendered.
The benchmark for avatar rendering (see Figure \ref{fig-2}) points the limit of 10 (Quest 1) or 23 (Quest 2) simultaneous avatars, but in real scenes, when buildings and particle effects are also rendered, the actual limit is probably lower. 
We also consider the IPD. Instead of a world established with fixed players at a fixed distance, as in a soccer game, Proxemics in Virtual Reality: What Should We Put to the Test in Social VR? (Welsch et al.)\cite{proxemics} argues the appropriate distance and IPD should be designed with culture in mind. We assume that various avatars communicate at a maximum distance of 1–2 meters, while generating collisions for individual avatars and avoiding overlap. Considering the interference in stereophonic sound, we consider that 5–11 avatars can be displayed simultaneously while maintaining quality communication.
After creating and testing a school life metaverse experience, we found that it is possible to provide users with a musical experience using a variety of avatars and effects if the number of simultaneous connections is approximately four (see part V/B.).

\begin{figure}[!ht]
\centerline{\includegraphics[width=1\linewidth]{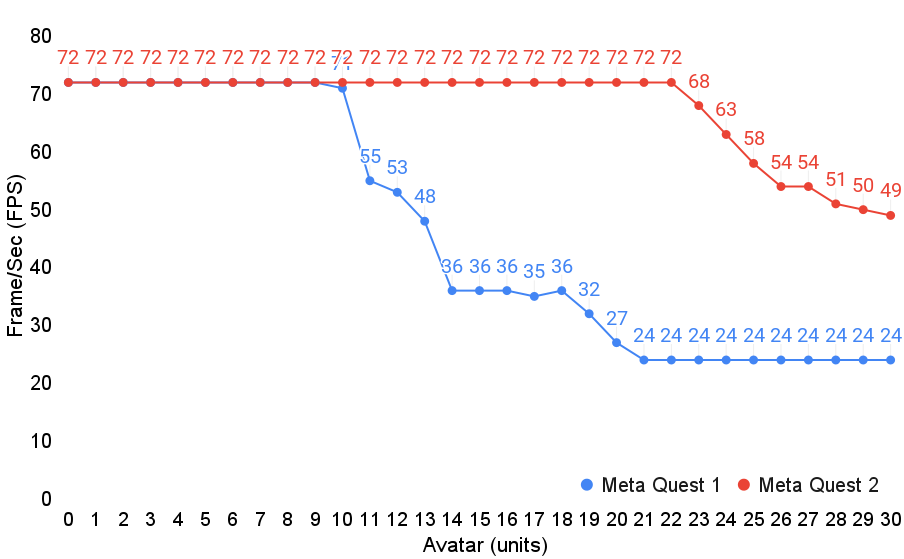}}
\caption{Graph showing the FPS depending on the number of avatars drawn on Meta Quest 1 and Meta Quest 2. On Meta Quest 1 the FPS starts dropping at 11 avatars, while on Meta Quest 2 it does not drop until 22.}
\label{fig-3}
\end{figure}

\subsection{Look dev and compatibility}
\begin{figure}[hbpt]
\centerline{\includegraphics[width=\linewidth]{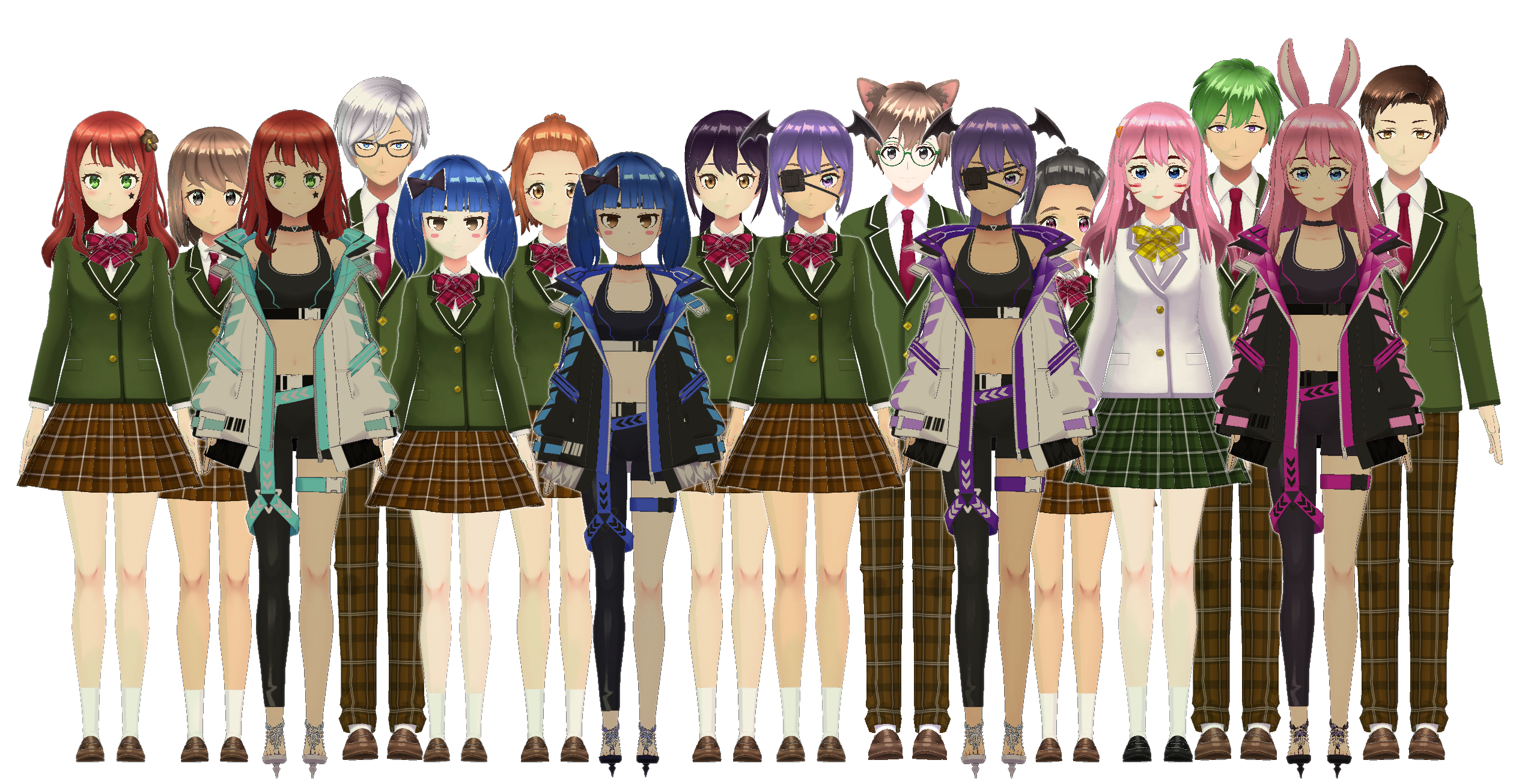}}
\caption{Example of different character models used in our application. The same character can wear a vast variety of clothes making the compatibility and look dev that more challenging.}
\label{fig-4}
\end{figure}
In a metaverse space, avatar materials, shaders, and room lighting must be set so that the avatar's appearance does not change significantly. Additionally, in a cross-platform metaverse where multiple platforms are connected, the avatar may need to be optimized for the user’s device environment. In such cases, it is important to consider atlasing the avatar's materials and textures (see Figure \ref{fig-4}), and maintaining compatibility with the physics applied to the hair and clothing.
We must consider the types of movements an avatar will make when deciding on its costume.
These could be fairly limited, such as walking, or extensive, such as dancing.
For instance, one potential issue is when the legs often penetrate the costume if the avatar is wearing a skirt (see Figure \ref{fig-4}).
We now know that our PoC should account for various limitations such as the performance, IPD, and look dev.

\section{AI-Assisted music play}

\subsection{Concept of AI Fusion}
\begin{figure}[htbp]
\centerline{\includegraphics[scale=0.1]{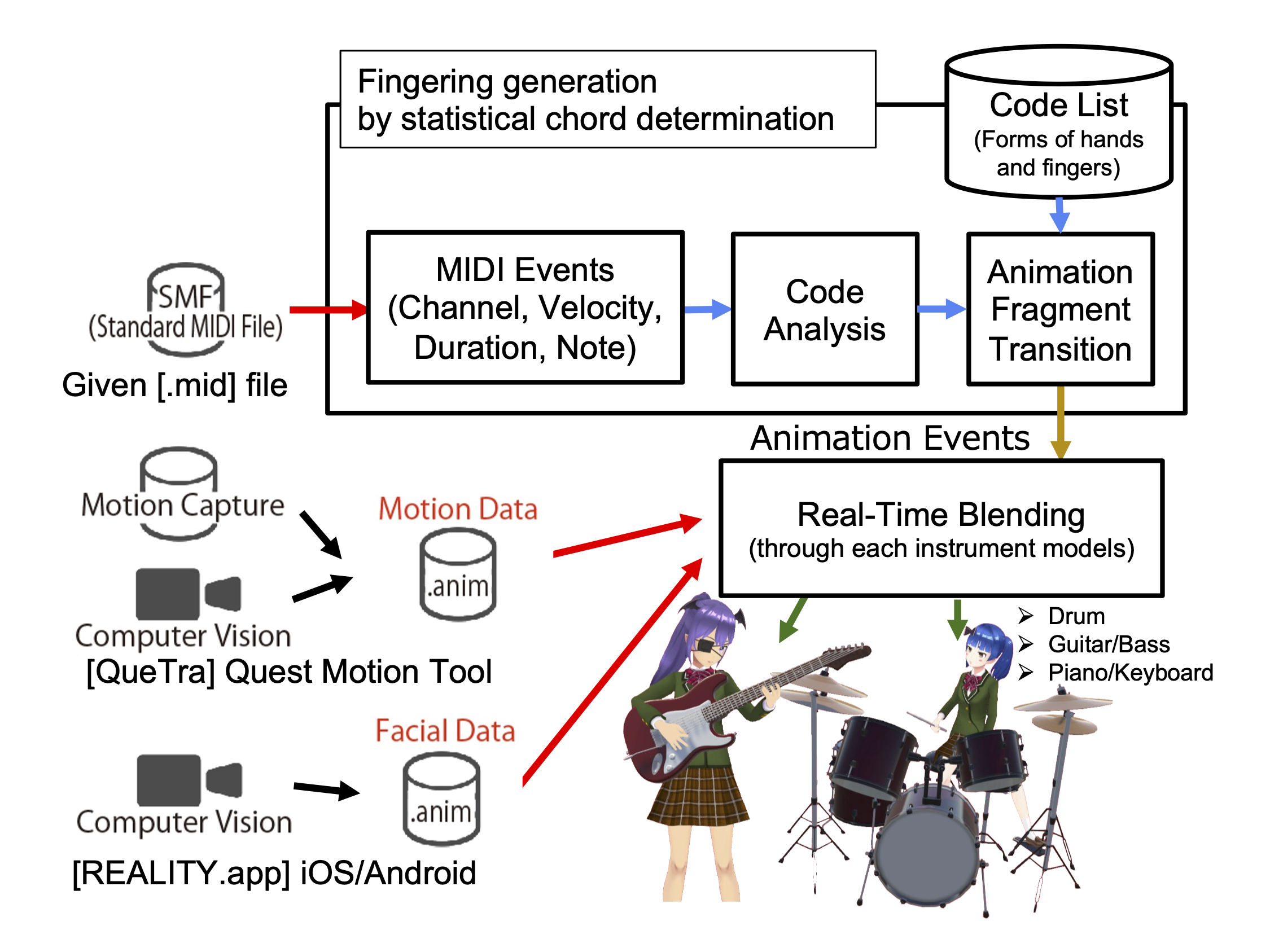}}
\caption{Diagram representing how the AI Fusion system works. We can simulate precise and natural music movements using MIDI files, motion-capture animation, and computer-created animation.}
\label{fig-5}
\end{figure}

To realize avatar karaoke, we created an automatic avatar music player system using SMF files (music scores, Standard MIDI, .mid) and AI Fusion (see Figure \ref{fig-5}). Similar attempts already existed, such as AI Concert Creator (https://www.concertcreator.ai/). The play is superb and precise. However, if the AICC example is watched in stop-motion, we observe ingenuity in piano playing, not only in the music but in the fingering and the rests. Musicality can generate these elements, such as in jazz or classical piano. As a metaverse service, they should be processed as the real-time movements of avatars in individual user performances (rather than being uniquely generated from scores or sound sources). Musical experiences in the metaverse are not only about playing instruments but also about singing. The technology for generating facial expressions from audio, especially mouth shape, has been available in the film and animation industry using LipSync for many years. It is available as a real-time animation generation technology in the Meta Oculus OVR LipSync. However, many basic technologies are based on vowel classification in speech signals. While the degree of mouth opening and closing and related shapes are classified, the technologies that change the degree of mouth opening and closing according to singing intensity are not generalized.
In addition, performers express themselves not only with their mouths when singing but also with their facial expressions and body movements. Considering the current use of stand-alone HMDs, the system targets photorealistic avatars and avatars with limited control rigs, such as in manga and anime. In the future, this could be controlled by machine learning, and we already know that a classifier for unique facial expressions is possible, as shown in AlterEcho: Loose Avatar-Streamer Coupling for Expressive VTubing (Tang et al.)\cite{AlterEcho}.
This study examines the generation of avatar facial expressions using recorded speech in multiple language environments while highlighting the limitations of expression in singing choruses. Can the results analyzed by audio files with NVIDIA Audio2Face be used for singing choruses? If they can, what are the potential limitations and possible solutions proposed?

\subsection{AI Fusion implementation in Unity}
As previously introduced, the AI Fusion system uses a mix of MIDI-generated and motion-capture animations. First, we examine how the MIDI-generated animations are handled in Unity.

\subsubsection{MIDI-generated animations}
\begin{figure}[!ht]
\centerline{\includegraphics[width=\linewidth]{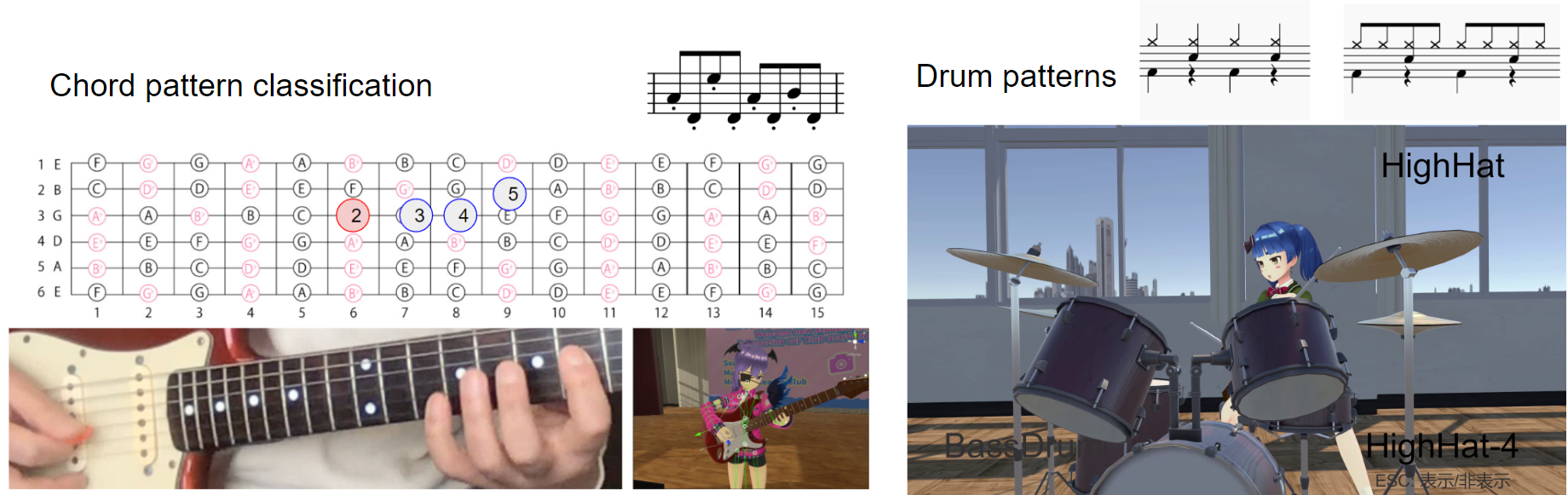}}
\caption{We classified the hand placement used for actual chords to build a classifier of an AI Fusion system. Drums can be expressed as single notes to classify a chunk of play for fluent animation, not only beats.}
\label{fig-6}
\end{figure}

For each instrument, we create animations to represent motions associated with each note we play. For every animation, we set the avatar position in one frame. For the bass or the guitar, we focused on the fingers’ positions to mimic movements (see Figure \ref{fig-6}).

\begin{figure}[!ht]
\centerline{\includegraphics[width=\linewidth]{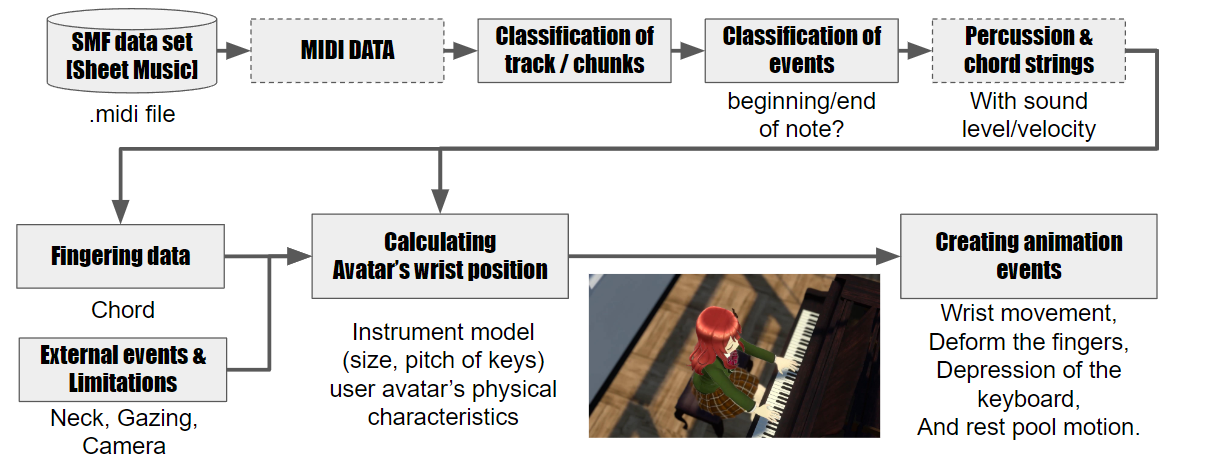}}

\caption{The AI Fusion system uses MIDI file inputs and converts them to Unity animation events, which we then use in our application in real time. The piano is a complex model with percussion, chords, and limitations on the avatar’s physical characteristics. Neck movement, gaze, and rest motions are essential to the player’s expression. The system can catch live expressions or gazes from other players or the audience.}
\label{fig-7}
\end{figure}

We used MIDI files as a base to link those animations to the music being played in real time (see Figure \ref{fig-7}). A MIDI file comprises chunks of data containing information about the tempo, note timing, velocity, and index. When the simulation starts, the MIDI file is read by a script that will retrieve those pieces of information from the MIDI file.

With the information we retrieved from the MIDI files, we know what note (index) is playing at what time (timing) and with which strength (velocity). To play these events with the proper timing, we used the Unity animation system. This is an easy way to play discrete events at a specific time. By creating a single clip in which we add animation events for every note, we automatically call a function every time a note is played in the MIDI file. Those animation events contain two pieces of information: the name of the function to call and the note code. When the clip is played, the character will recover the animation events and dynamically change its animation using the preemptively made animations. This system reached a high level of movement reproduction fidelity. Yet the character felt static and robotic even though it made no mistakes and its movements were accurate. To compensate, we mixed these animations with motion-capture data.

\subsubsection{Motion capture animations}
We went to a specialized studio to capture animation samples that we could use in Unity. 
Those motion-capture samples were sent as objects from which we could create .anim files and apply them to any character. 

\begin{figure}[!ht]
\centerline{\includegraphics[scale=0.45]{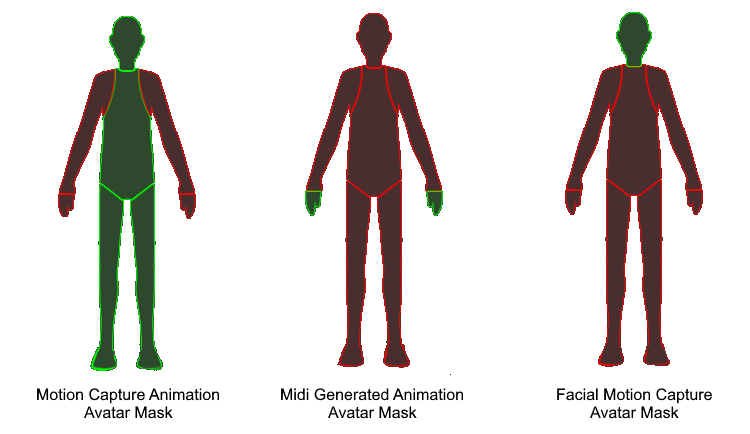}}
\caption{For each avatar we control which body part will be controlled by which animations using a mask system. For the bass, only the arms are controlled by computer-created animation.}
\label{fig-8}
\end{figure}

To mix the animations from the motion-capture samples, the facial captures, and the MIDI-generated animation, we used the animation layer system from Unity. Using the animation layer system, it became possible to apply animations only to certain parts of the avatar’s body, which are defined by an avatar mask (see Figure \ref{fig-8}). It is also possible to apply animation to the same body part as ``additive.'' By doing so, the head movement is created from the motion-capture animations,
and the facial expressions are created from the face-motion capture.

\subsubsection{Expression as a short film}
\begin{figure}[!ht]
\centerline{\includegraphics[scale=0.17]{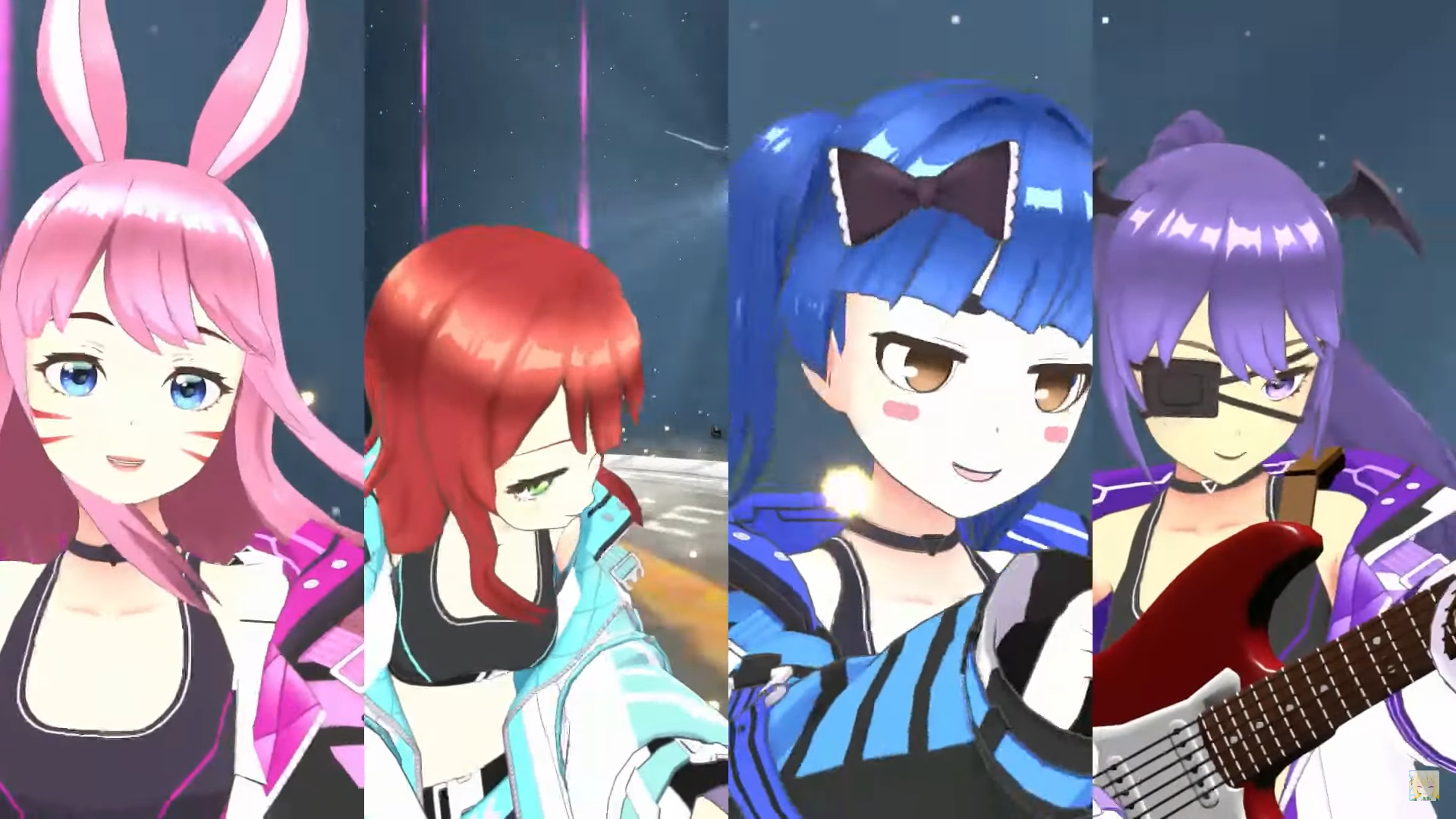}}
\caption{We posted the short film on our team’s YouTube channel, where it has approximately 2,200 views.  \url{https://youtu.be/bU8QUlxOTFU}}
\label{fig-9}
\end{figure}

To showcase the AI Fusion system, we made a short film, MetaDreamers (see Figure \ref{fig-9}). Our objective was to test the AI Fusion system and, most importantly, its limits, and iterate and improve it. In the clip, we see four different instruments: a guitar, a bass, a piano, and drums. By playing multiple instruments, we had the opportunity to challenge the AI Fusion system with different types of animation. We iterated on the AI Fusion system during production to resolve specific issues by blending the animations (for example). Some problems remained, such as the wide and irregular movements of the piano and drums.

One of the challenges of using the AI Fusion system was finding which animations control which parts of the avatar. Only the fingers use the MIDI-generated animations for the guitar and bass, while the motion-capture data control the rest of the body. While this works well for the guitar and bass, the results for the piano and drums are more mixed.

Ideally,  the drums can be played perfectly as an autonomous machine, but it looks less natural(\url{https://youtu.be/LrB5aNikLvA}).

AI Fusion can express the head, neck, hair and face. Only the head, neck and hair are not constrained, and the precessions and rests are important elements in the expression. The piano is both a percussion and a string instrument. The playing of music constrains the arms, hands and feet.

The head movement, eye gaze, and mouth position can give various impressions, and MIDI-generated animations and the large-space motion-capture data do not produce them. Our smartphone-based capture system complements human expressions. We achieved more consistent results with better weight distribution between the avatar masks. It could be possible to dynamically change some animations depending on the current avatar position or muscle force. The system could control visual effects. Indeed, the AI Fusion system offers enough information to launch effects synchronized with the music. By using the velocity parameter of the MIDI files, it could be possible to control the intensity of those effects.

Our understanding and the challenges of advanced human-machine collaborative instrumentation and expression have contributed to the deepening, generalization, and theorization of AI Fusion technology. We believe this technology can be applied to other areas (such as walking and dancing).

\subsection{Facial animation and voice recognition}

\begin{figure}[!ht]
\centerline{\includegraphics[width=\linewidth]{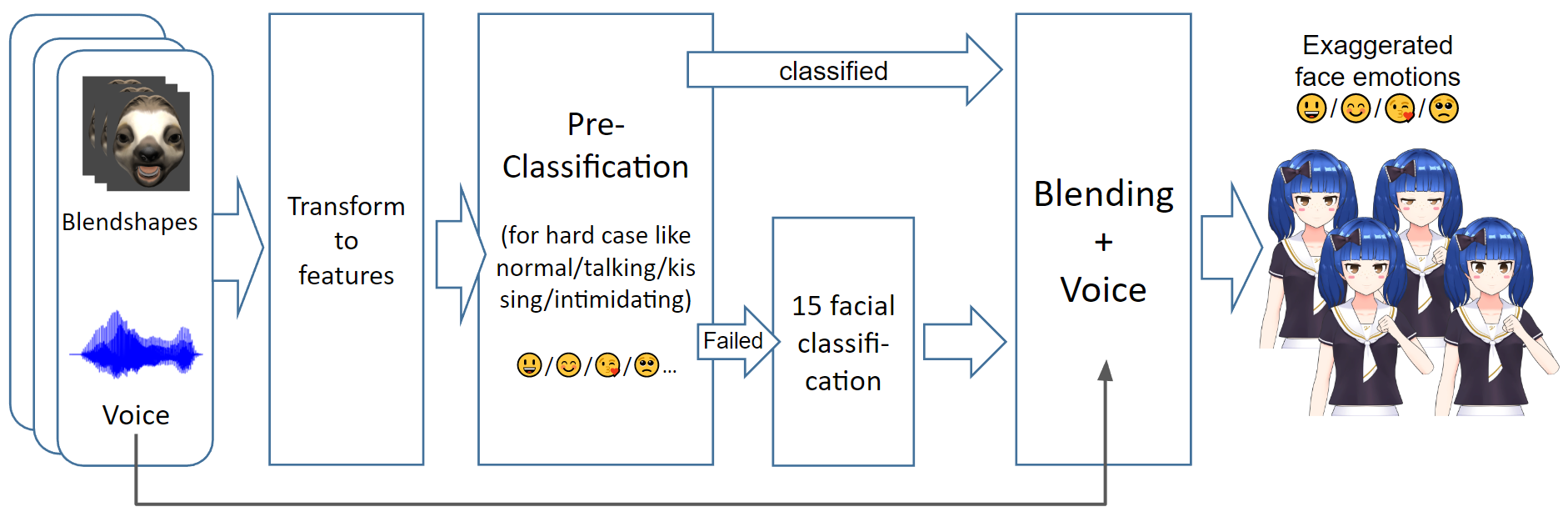}}

\caption{The process to generate exaggerated facial expressions with Blend- Shape + LipSync from real-time voice and depth-camera input.}
\label{fig-10}
\end{figure}

For avatar karaoke in VR to be successful, the avatars must make different facial expressions in sync with their singing. They must easily convey emotions, and because the metaverse is used globally, these emotions need to be shared among people from different cultures and languages. This can be achieved using the international language of emoticons; they do not require translation and can be understood by everyone. The exaggerated faces of avatars are similar to emoticons. Most animated avatar expressions, such as those of VTubers’ avatars, are animated using facial recognition from a camera. However, using an HMD does not allow for accurate facial recognition. Therefore, we need to investigate voice and speech recognition to create synchronized lip sync and facial animation. Also, tracking data cannot be applied directly to exaggerated faces. We explored various options, including rule-based state machines and deep-learning base classifiers (see Figure \ref{fig-10}), using random forest, the depth camera of an iPhone X, and blendshapes. The result is a vision system for faster smartphone tracking, with frame rates of 30 fps or more and fewer control points for a more realistic representation. In other words, high-frame-rate feedback loops were found to handle diverse user-intended representations better than complex decision algorithms.

\begin{figure}[!ht]
\centerline{\includegraphics[scale=0.3]{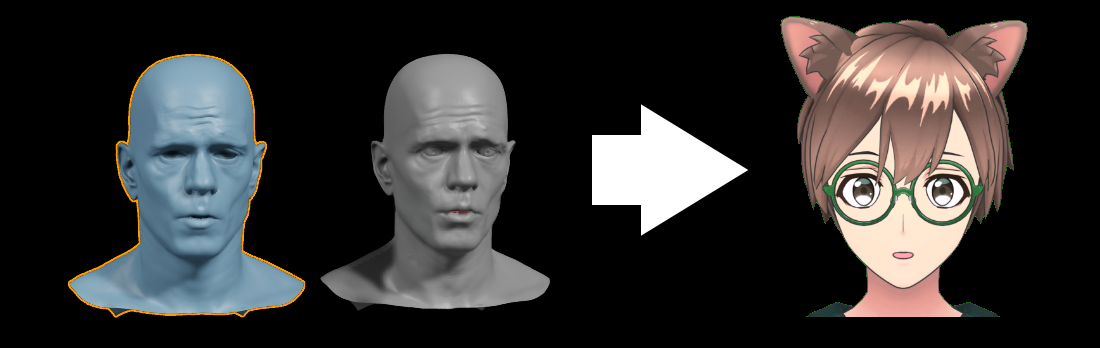}}
\caption{Using only a part of the values generated by Audio2Face, we created a new facial animation pipeline on a VRM avatar.}
\label{fig-11}
\end{figure}

Another option was to use NVIDIA’s Audio2Face application. In theory, this application can simulate a wide variety of facial expressions and emotions by using only voice input. However, we quickly understood that it would be difficult to implement the complexity of all the parameters that Audio2Face manages onto anime-type avatars, as they have far fewer parameters available to modify (see Figure \ref{fig-11}). At the time this paper was written, Audio2Face supports only some Latin-based languages such as English, French, or Spanish. Japanese, Chinese, or Korean are not supported, meaning that the lip-syncing is not accurate. Lastly, we have yet to test Audio2Face's real-time lip sync in an application with Audio2Face.

Tang et al. \cite{AlterEcho} propose a solution to dynamically play animations using various sources of input such as voice, facial capture from ARKit, and personality factors. In this case, speech recognition cannot control facial expressions directly; it triggers premade animations. While the system can work in a chat scenario, it cannot convey the emotion and intensity of a singing session.

In this research, we evaluated the current state of the art of various AI-driven animation technologies, what these technologies can create, and the related issues that still need resolution.

\section{Tools and Proof of concepts}

\subsection{QueTra: Simple Motion Capture system using Quest}
\begin{figure}[!ht]
\centerline{\includegraphics[scale=0.4]{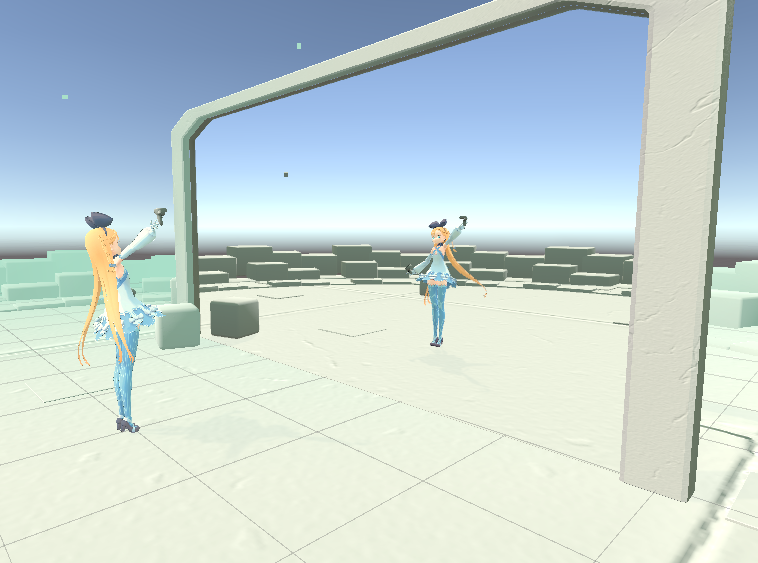}}
\caption{In-application screenshot of QueTra. The user has a mirror to see the animations easily.}
\label{fig-12}
\end{figure}

QueTra is a simple motion capture system that uses Meta Quest to create avatar animations compatible with Unity’s humanoid format. It supports animation output in BVH format, making it suitable for a range of uses. In addition to production for internal projects, the system was used in a children’s workshop, where participants created their own metaverse. Children made avatars with dance motions in Unity using QueTra, which they ran on Web3D and viewed from their smartphones (see Figure \ref{fig-12}). 
The initiative to turn Quest into a simple cloud-based motion-capture system has various potential applications, although Quest does not have an interface to a PC. Its development will continue in cooperation with AI Fusion techniques.

\subsection{GVBand: a prototype of a multi-user music-playing VR experience in the metaverse}
\begin{figure}[!ht]
\centerline{\includegraphics[scale=0.04]{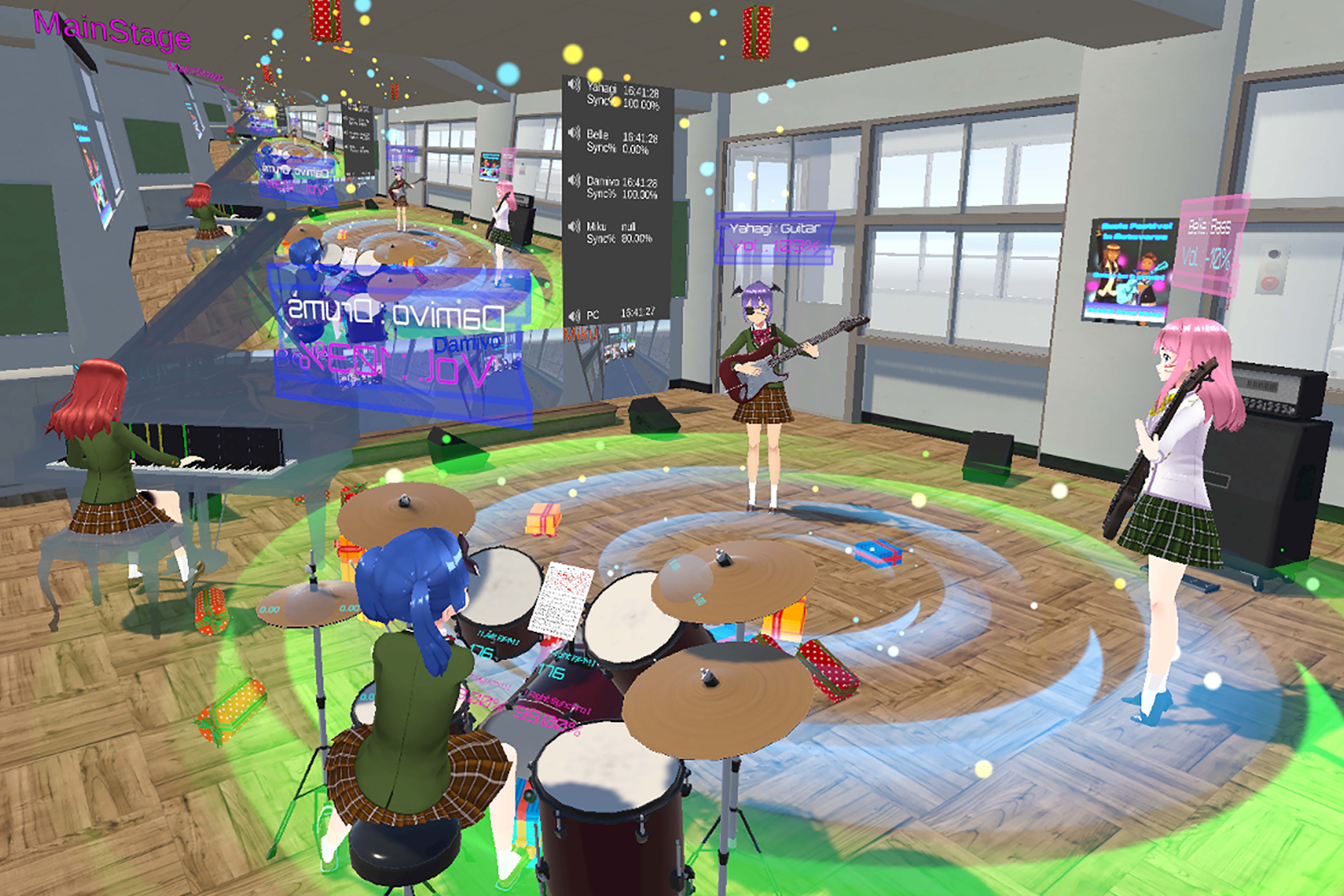}}
\caption{GVBand PC instance to observe HMD players on the network.}
\label{fig-13}
\end{figure}

In the GVBand application (see Figure \ref{fig-13}), the user enters a metaverse and practices for a performance as a high-school music club member. Up to four users can play at the same time. No guitar, bass, drums, or piano knowledge is required. As participants play along with the music, their “excitement level,” which is visualized as a numerical value, increases. As the excitement level rises, effects are displayed, and gifts appear. Up to four independent Quests can be in session simultaneously, including voice chat via the cloud. Using AI Fusion, the neck, head, and one hand can be used for expression while playing an instrument. This may not sound like much movement, but as mentioned previously, we found that individual musical expression includes gestures of the neck and the direction of the gaze and occurs during rests and intervals in the performance and in the consensus building among individual users.

\section{Conclusion and future work}
\begin{figure*}[ht]
\centerline{\includegraphics[width=0.85\linewidth]{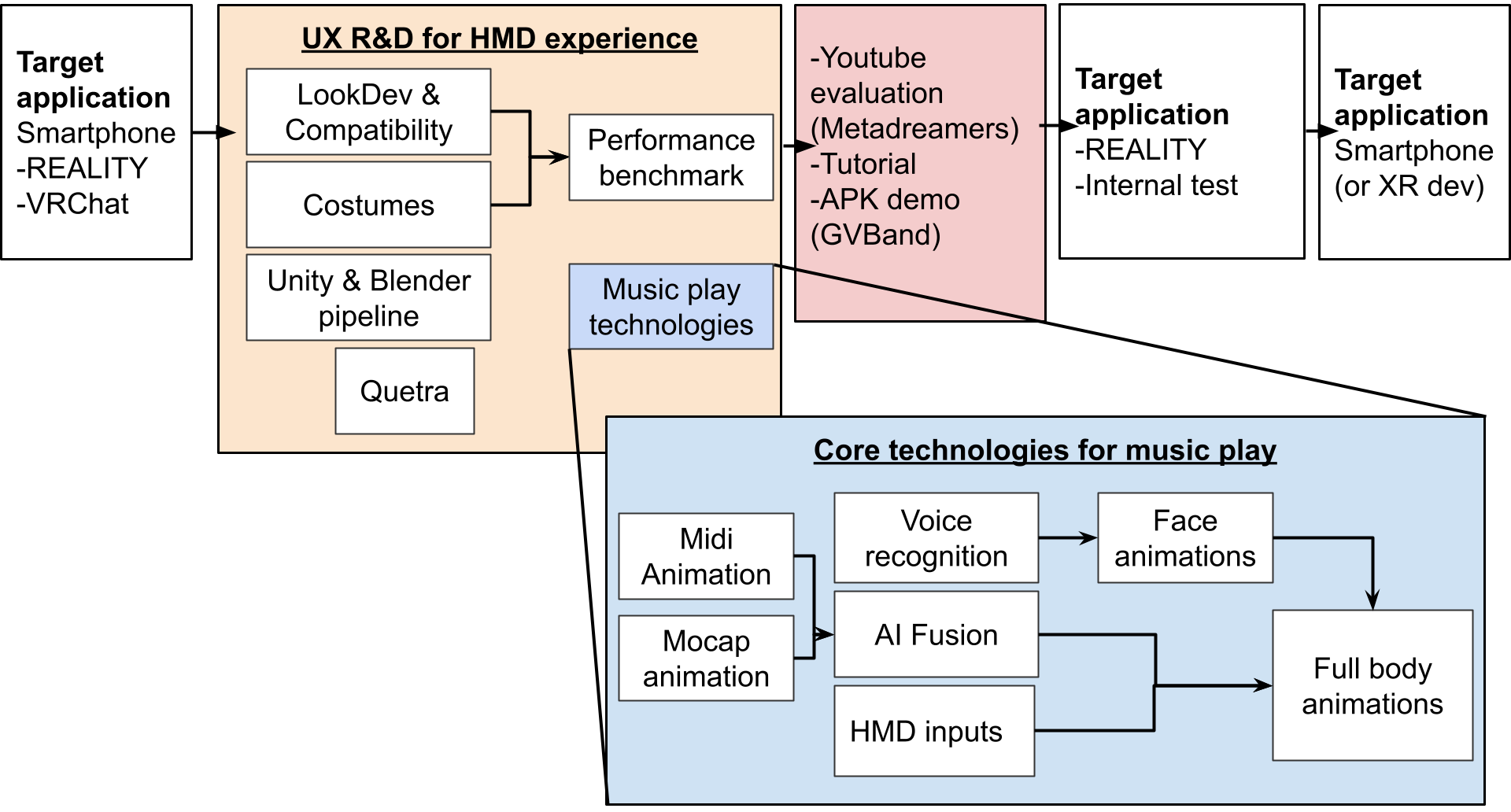}}
\caption{We started with the objective of an avatar-driven music play. Through various R\&D, we created an APK application (GVBand) and a short film (MetaDreamers), which can then be evaluated.}
\label{fig-15}
\end{figure*}

As we hypothesized at the beginning of this paper, we developed a PoC for metaverse R\&D that could be evaluated internally and by users. We generated a variety of content and UGC using AI Fusion avatar karaoke. We now foresee the volume of technical issues and work hours required to develop such content, including the use and licensing of tools and MIDI scores, the use of user tracking information,
and the development of art assets, such as animation elements (see Figure \ref{fig-15}). We have reported on several PoCs based on the idea that we need to create an atavistic experience on HMDs from the avatar-driven SNS, which can be categorized into the following three types of validation. First, we need to guarantee the quality of communication based on the limitations of the devices and 3D human vision. It was necessary to use current devices to start with and obtain benchmarks for simultaneous avatar drawing in Meta Quest 1 and Meta Quest 2. Even if future performance improvements increase the number of simultaneous renders, it is not reasonable to assume that all users will simultaneously use state-of-the-art hardware for a service. Detailed benchmarking revealed that the simultaneous rendering of avatars of the quality currently running on smartphones is 10 for Quest 1 and 23 for Quest 2. Second, we identify issues and challenges in the user collaboration experience that must be addressed. Unlike a gamelike metaverse based on artificial, prearranged rules, VR SNS must allow spontaneous user interactions. The school-life metaverse was a suitable subject for discovering problems and challenges in these simultaneous connections. To qualitatively evaluate the potential for a multi-person metaverse experience, this study examined subject matter that required each avatar to be expressive. The need to design for appropriate IPD and account for cultural differences will remain unchanged. Specifically, individual avatars should collide (if there are no collisions, the interior of others’ avatars will be drawn, which is more costly), a way to move while avoiding overlap, consider stereophonic interference in voice communication, speak at a maximum of 1–2 meters, and display 5–11 avatars simultaneously. A space displaying 5–11 avatars simultaneously would be an appropriate environment in which to maintain communication. Finally, we need to verify user motion representation and complementary technologies. Suppose it is not possible to create an environment where all users stream motion data in real time. In that case, it is necessary to have playback motion with automatic generation and completion that is symbolized to some extent. We developed a scene using QueTra and AI Fusion to give meaning and expression to the user’s motions by guaranteeing accuracy and quality (e.g., real-time performance of a song) and used this as a fundamental technology for future research. An example implementation in Unity is presented along with this paper. Through these three PoC validations, we created and experimented with a prototype of Avatar Fusion Karaoke and reported the following qualitative effects we did not anticipate before the development. 
The same session showed significant effects with multiple participants. Most users did not recognize the AI-only piano-playing avatar as a human (i.e., they perceived it as an NPC). When users waved to one another (saluted), their participation time increased by about 3 minutes (i.e., they played more than one song). From these results, we can say that AI Fusion is more effective than bot users. Although this study has been conducted in a school setting, it can be applied to corporate activities, such as shopping malls and conferences, for comfort, quality assurance, and optimization of collective behavior. It can also be applied to real-life education. 
For example, it could be applied to study students' influence on one another's behavior during remote learning using videoconferencing. 
We started with the objective of testing a standard R\&D process for a new metaverse UX.
To this end, we created an avatar-driven music-playing experience to illustrate some valuable scenes for a multi-user 3D avatar metaverse. Through various R\&D efforts, we developed QueTra, a motion capture tool for a stand-alone HMD, a cloud HMD communication application (GVBand), and a short film (MetaDreamers) for evaluation. We have not previously recognized the PoC and the development of stand-alone technologies. Still, these expressive technologies are important in a multi-user context precisely because they are directly related to the quality of communication within the metaverse and the value of the UGC produced there and should be fully recognized. We could propose a PoC and public demonstration of Avatar Fusion Karaoke, as it was confirmed to be a valuable application. To standardize this PoC, we will reflect on current communication karaoke-related technologies. As multi-user music-playing VR experiences in the metaverse will be enhanced in the near future, AI-assisted technologies, including motion fusion, completion, and compression techniques, will also see enhancements.


\printbibliography

\end{document}